\documentclass[amsmath,amssymb,amsfonts,showpacs,floatfix,twocolumn,pra,superscriptaddress]{revtex4}
\usepackage{graphicx}
\usepackage{dcolumn}
\usepackage{bm}
\usepackage{times}
\usepackage{color}

\newcommand{\bk}{\mathbf{k}}
\newcommand{\bq}{\mathbf{q}}
\newcommand{\br}{\mathbf{r}}
\newcommand{\la}{\langle}
\newcommand{\ra}{\rangle}
\newcommand{\ket}[1]{|{#1}\rangle}
\newcommand{\bra}[1]{\langle{#1}|}

\newcommand{\norm}[1]{\ensuremath{| #1 |}}

\newcommand{\rr}{\mathbf{r}}

\newcommand{\eqname}[1]{\label{eq:#1}}
\newcommand{\eq}[1]{(\ref{eq:#1})}
\newcommand{\psih}{\hat{\psi}}
\newcommand{\psihd}{\hat{\psi}^\dagger}

\begin{document}

\title{All-optical pump-and-probe detection of two-time correlations in
a Fermi gas }

\author{T.-L Dao}
\affiliation{Centre de Physique Th{\'e}orique, {\'E}cole
Polytechnique, CNRS, 91128 Palaiseau, France.}

\affiliation{Laboratoire Charles Fabry de l'Institut d'Optique,
CNRS and Univ. Paris-Sud, Campus Polytechnique, RD 128, F-91127
Palaiseau cedex, France.}

\author{C. Kollath}
\affiliation{Centre de Physique Th{\'e}orique, {\'E}cole
Polytechnique, CNRS, 91128 Palaiseau, France.}

\author{I. Carusotto}
\affiliation{CNR-INFM BEC Center and Dipartimento di Fisica, Universit\`a di Trento, 38050
Povo, Italy}
\author{M. K\"ohl}
\affiliation{Cavendish Laboratory, University of Cambridge, JJ
Thomson Avenue, Cambridge CB3 0HE, United Kingdom}

\begin{abstract}
We propose an all-optical scheme to probe the dynamical correlations of a strongly-interacting gas of ultracold atoms in an optical lattice potential.
The proposed technique is based on a pump-and-probe scheme: a coherent light pulse is initially converted into an atomic coherence and later retrieved after a variable storage time.
The efficiency of the proposed method to measure the two-time one-particle Green function of the gas is validated by numerical and analytical calculations of the expected signal for the two cases of a normal Fermi gas and a BCS superfluid state.
Protocols to extract the superfluid gap and the full quasi-particle dispersions are discussed.
\end{abstract}

\date{\today}

\maketitle

\section{Introduction}
Many-body quantum systems exhibit truly
remarkable features such as high-temperature superconductivity and
the fractional quantum Hall effect. Traditionally, these phenomena
are studied in the solid state. However, in recent years dilute,
yet strongly interacting, atomic gases have started providing a
novel class of systems to investigate this fascinating physics.
Their outstanding cleanliness, control, and precise microscopic
understanding will push forward the fundamental understanding of
quantum many-body physics \cite{Bloch2008}.

Strongly interacting atomic quantum gases are generally prepared
by trapping atoms in vacuum in a magnetic or optical potential.
This offers two remarkable opportunities: Firstly, a superb
isolation from the environment opens the door to fascinating
experiments out of equilibrium to investigate genuine quantum
dynamics.
 Secondly, a variety of coherent optical processes are
available to selectively probe the quantum system without being
disturbed by a surrounding bulk medium. In particular, these
optical detection techniques can provide repetitive and almost
non-destructive in-situ measurements
\cite{Andrews1996,Higbie2005}. The combination of these two
features makes ultracold quantum gases ideal systems to study the
non-equilibrium and dynamic properties of isolated quantum many
body systems. However, the experimental study of these properties requires the
development of novel detection schemes that are sensitive to a wider
variety of observables of the quantum gas, e.g. its multi-time correlation
functions.

The prime example of an atomic quantum system mimicking the
physics of the solid state are interacting fermionic atoms in
artificial lattices structures, the so called optical lattices
\cite{Koehl2005}. The preparation of strongly correlated states in
an optical lattice will allow for an analog simulation of complex
quantum many body Hamiltonians. Recently, evidence for the
stabilization of a Mott-insulating phase has been obtained by
looking at density related quantities of the gas
\cite{Joerdens2008,DeLeo2008,Schneider2008,Scarola2009}.
The
identification and characterization of more complex quantum phases
requires, however, the measurement of time-resolved
single-particle correlation functions, also called Green functions,
of the form $\langle
\psi^{\dag}_{\sigma,r}(t)\psi_{\sigma',r^\prime}(t^\prime)\rangle$.
Here $\psi^{(\dag)}_{\sigma,r}(t)$ is the annihilation (creation)
operator for the internal atomic state $\sigma$ at position $r$
and time $t$. The single-particle two-time correlation function reveals
profound information about the macroscopic coherence and
decoherence of the systems and keeps track of the subtle properties of quantum
 phases which are not density-ordered, e.g. the existence of quasi-particles in a strongly correlated Fermi liquid.  This same correlation function plays a even more crucial role in the case non-equilibrium situations: as the most celebrated example, the particular relaxation behaviour of glasses is almost invisible in one-time correlations, while it can be followed in full detail by measuring the two-time ones \cite{Cugliandolo2002}.

Up to now, the single particle equal time correlation function out of equilibrium was investigated for bosonic atoms \cite{Ritter2007}.
For fermions, only the energy resolved correlation function of an equilibrium state has so far been probed
by momentum-integrated \cite{Toermae2000,Regal2003,Gupta2003,BruunBaym2004,Chin2004,Jiang2009} and
momentum resolved \cite{Dao2007,Stewart2008} radio-frequency or two-photon spectroscopy.
A more elaborated scheme for the detection of the two-time correlation function based on the immersion of a ion into quantum gases has been proposed in \cite{Kollath2007}.

Here, we propose an all-optical pump-and-probe scheme to extract
quantitative information on the microscopic physics of a Fermi
gas and in particular on its two-time correlation functions.
A pump sequence firstly brings the system into a quantum
superposition of its initial state and an excited state.
The response of the system to a second probe pulse sequence is then
measured after a variable time delay. In this way, information on
the time evolution of the atomic two-time correlations is converted into
easily detectable observables, such as the intensity and the phase
of the outgoing light.

From an alternative point of view, our scheme can be seen as an
application of light storage techniques
\cite{Liu2001,Schnorrberger2009,Hau_Nature2007} to the diagnostic
of many-body systems: a coherent pulse of light is stored in a
quantum gas and retrieved at a later time after a variable
interval. Information on the system is extracted from the
properties of the retrieved light. Differently from standard light
storage experiments where it is a purely detrimental effect,
decoherence of the stored pulse as a function of storage time is
in our scheme the crucial tool to obtain information on the
many-body dynamics of the underlying quantum gas.

The structure of the paper is the following. In Sec.\ref{sec:scheme}
the measurement schemes are introduced and analytical expression relating
the observed signal to the many-body observables are given. In Sec.\ref{sec:BCS}
an application to a fermionic system is discussed in detail and experimental protocols
to extract the superfluid gap and the full quasi-particle dispersions of a BCS superfluid
are outlined. 
Conclusions and future perspectives are given in
Sec.\ref{sec:conclu}.

\section{The measurement procedure}
\label{sec:scheme}
\begin{figure}[!ht]
    \includegraphics[width=.5\columnwidth,clip=true]{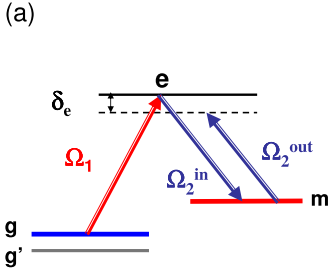}
  \caption{(Color online) Diagram of the internal atomic levels involved in the proposed detection scheme.
  } \label{fig:schemea}
\end{figure}

\begin{figure}[!ht]
    \includegraphics[width=0.9\columnwidth,clip=true]{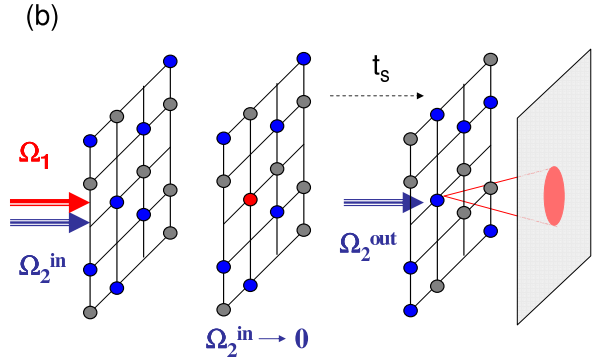}
  \caption{(Color online) Snapshots of the measurement procedure. From left to right: adiabatic storage of coherence from incident beams, free many-body evolution, light re-emission and detection.
  } \label{fig:schemeb}
\end{figure}

Even though the application scope of our measurement procedure is much wider, we focus our attention here onto the case of fermionic atoms in a two-dimensional lattice geometry. This geometry lies at the heart
of quantum simulations with the aim of exploring the mechanisms
underlying high-temperature superconductivity
\cite{Hofstetter2002}. A tight optical confinement potential
freezes the atomic motion into a single $xy$ plane. Additionally,
a periodic optical lattice potential is applied along the $x$ and
$y$ directions to generate a two-dimensional lattice structure
\cite{Gemelke2009}.

The gas consists of a mixture of atoms in two
hyperfine ground states $g$ and $g'$ that feel an identical
confinement potential. Our all-optical probing scheme involves
three atomic levels ($g,e$, and $m$) arranged in a $\Lambda$
scheme as schematically shown in Fig.~\ref{fig:schemea}. The $m$
state is a long-lived electronic ground state whereas the $e$
state is an electronically excited state. With a suitable choice
of polarization and frequency, the $g'$ atoms experience a
negligible coupling to the pump and probe light fields.

The diagnostic scheme (Fig.~\ref{fig:schemeb}) starts with the
creation of a coherent excitation by adiabatically
switching on a laser of (spatially uniform) Rabi frequency
$\Omega^{\rm in}_2$ and then a weaker collinear laser of
(spatially dependent) Rabi frequency $\Omega_1(\rr)$
\cite{Fleischhauer2005}. The two beams are then suddenly and
simultaneously switched-off. The frequency $\omega_{1,2}$ of the
 beams are chosen to be resonant with the $g
\rightarrow e$ and $m \rightarrow e$ transitions, respectively. To
ensure adiabaticity of the preparation stage, the switch-on of the
two lasers has to be performed on a time-scale long as compared to
the internal atomic dynamics and to the Rabi frequencies
$\Omega_1$, $\Omega_2^{\rm in}$. On the other hand, the switch-off
has to be much faster than all these frequencies.

Provided the whole excitation stage takes place on a fast time
scale as compared to the many-body dynamics, the atomic position
can be considered as fixed and the optical process described by a
single atom picture where the atomic $|g\ra$ state is
adiabatically transformed into a dark state
\cite{Dutton2004,Fleischhauer2000,Juzeliunas2004,Carusotto2008}
\begin{equation}
\ket{\textrm{dark}}=\frac{e^{i \theta}}{\sqrt{\norm{\Omega_1}^2+\norm{\Omega_2^{in}}^2}}\,\left[\Omega_2^{in}\ket{g}-\Omega_1\ket{m}\right]
\end{equation}
which is fully decoupled from the excitation lasers.
All other bright eigenstates are energetically separated and do not get mixed with the dark state provided the switch-on phase is performed in a slow enough manner.
Assuming that the phase of the $\Omega_{1,2}$ Rabi frequencies (once the carriers at $\omega_{1,2}$ are eliminated) is constant during the whole sequence, the global Berry phase
\begin{equation}
\theta=i\!\int\!dt\,\la\textrm{dark}(t)|\frac{d}{dt}|\textrm{dark}(t)\ra
\end{equation}
acquired by the atom during the adiabatic evolution is easily shown to vanish.

In the $|\Omega_1|\ll |\Omega^{\rm in}_2|$ limit under
investigation here, the effect of the excitation stage on the
initial many-body state $|\phi_0\ra$ in the lattice representation can be expressed as the
transformation:
\begin{equation}
\label{eq:dark}
|\phi_{d}\ra \simeq
\Big(\mathbf{1}-\sum_{i}\frac{\Omega_1(\br_i)}{\Omega_2^{\rm
in}}\, \psih^{\dagger}_{m,\br_i}\psih_{g,\br_i}\Big)|\phi_{0}\ra.
\end{equation}
Here, the $\psih_{\sigma,\br_i}$ ($\psihd_{\sigma,\br_i}$) lattice
operators destroy (create) a fermionic atom in the $\sigma=g,m$
state at the lattice site $\br_i$, respectively.  $\Omega_{1,2}$ are the Rabi frequencies in the lattice representation. Initially, only
the $g$ and $g'$ states are assumed to be occupied.

The use of a locally focused laser of amplitude $\Omega_1(\rr)$ on
the $g\rightarrow e$ transition allows one to selectively address
a well defined region of the sample. The space selection is useful
to eliminate inhomogeneous broadening effects due to the spatially
varying density in e.g. trapped systems
\cite{Yavuz2007,Gorshkov2008}. If no spatial selection is
performed, the final signal would include the contributions of
different regions of the system.

Once the two beams are switched off, the system evolves according to its many-body
Hamiltonian for a {\em storage} time $t_s$ from the state $|\phi_d\ra$
to the new many-body state
\begin{equation}
\ket{\phi(t_s)}= U_{mb}(t_s)\,\ket{\phi_d}=e^{-i\,(H_0+H_m+H_{\mathrm{int}})\,t_s/\hbar}\,\ket{\phi_d}
\label{eq:evol}
\end{equation}
where $U_{mb}(t_s)$ is the many body time-evolution operator and
the system Hamiltonian involves three contributions: $H_0$ is the
Hamiltonian acting on the states $g$ and $g'$, $H_m$ the
Hamiltonian of the atoms in the $m$ state, and $H_{\mathrm{int}}$
contains the interaction processes between the $g,g'$ and $m$
states. This time-evolution will change the coherence between $g$
and $m$ present in the prepared dark state. The $g,m$ coherence
remaining at the end of the storage time is finally probed.

The detection of the remaining coherence can be achieved by
different schemes: Either a fast $\pi$ pulse is applied to
coherently transfer all atoms from the $m$ to the $e$ state and
then coherent photons emitted on the $e\rightarrow g$ transition
are detected. Or the excitation is slowly released by means of a
weak field of frequency $\omega_2$ and Rabi frequency in the continuum
$\Omega_2^{\rm out}\ll \gamma_e$ that transfers the atoms
adiabatically from the $m$ state into a coherent superposition of
$m$ and $e$. In both cases, the electric dipole that is
responsible for the emission at frequency $\omega_1$ on the
$e\rightarrow g$ transition is proportional to the coherence
between the $g$ and $m$ atomic states,
\begin{equation}
\hat{d}(\br)=D\,\psih^{\dagger}_{g,\br}\,\psih_{m,\br}.
\end{equation}
The constant $D$ depends on the details of the process and
determines the duration in time $\tau_r$ of the released pulse
$\tau_{r}^{-1}=\gamma_e\,|d_{eg}/D|^2$, where $\gamma_e$ is the
radiative decay rate of $e$ state atoms. In the $\pi$ pulse case,
the constant $D$ is equal to the electric dipole matrix element
between the state $g$ and $e$, $D=d_{ge}$. In the case of a slow
release, $D$ is approximately given by $D=2i \Omega^{\rm out}_2
d_{ge}/\gamma_e$. In order for the many-body dynamics not to
interfere with the release process, the time duration $\tau_r$ of
this latter has to be shorter than the characteristic time scales
of the many-body dynamics.

The near field pattern of the emitted light {\em amplitude} is
determined by the expectation value of the local dipole operator
on the final state $\ket{\phi(t_s)}$
\begin{eqnarray}
d({\br},t_s) &&= \bra{\phi(t_s)} \hat{d}({\br})\ket{\phi(t_s)}
\label{eq:final_d}
\end{eqnarray}

We switch now to the lattice representation by relating the field
operators in the continuum to the lattice operators via the Wannier functions
$w_{\sigma}(\br)$ as following $\psih_{\sigma,\br}=\sum_i
w_{\sigma}(\br-\br_i)\psih_{\sigma,\br_i}$. We checked numerically that for tight atomic
Wannier functions on deep lattices, we can keep only the terms with Wannier factors taken at the same sites and neglect all other contributions. 
This
leads to the following expression for the dipole operator
\begin{eqnarray}
d({\br},t_s)
&&\simeq
D\sum_{i}W_{\br_i}(\br) \bra{\phi(t_s)}
\,\psih^{\dagger}_{g,\br_i}\,\psih_{m,\br_i}
\ket{\phi(t_s)}, \label{eq:final_d}
\end{eqnarray}
with $W_{\br_i}(\br)=  w^*_g(\br-\br_i)w_m(\br-\br_i)$.
Inserting the expression
(\ref{eq:evol}) of the final state and switching to the Heisenberg
representation for the operators $\psih_{x,\br_i}(t_s)=
U^\dagger_{mb}(t_s)\psih_{x,\br_i} U_{mb}(t_s)$, this has the form
\begin{eqnarray}
d({\br},t_s)= &&D\,\sum_{i}W_{\br_i}(\br) \nonumber \\
&&\times \bra{\phi_d} U^\dagger_{mb}(t_s)\,\psih^{\dagger}_{g,\br_i}\,\psih_{m,\br_i} U_{mb}(t_s) \ket{\phi_d}\nonumber \\
=D&&\sum_{i}W_{\br_i}(\br) \bra{\phi_d}\,\psih^{\dagger}_{g,\br_i}(t_s)\,\psih_{m,\br_i}(t_s) \ket{\phi_d}.
\label{eq:final_d}
\end{eqnarray}

Inserting into (\ref{eq:final_d}) the explicit expression (\ref{eq:dark}) for the dark state and taking into account that no atoms were initially present in the $m$ state, this expression can be written in the compact form
\begin{multline}
d({\br},t_s) = -\frac{D}{\Omega_2^{\rm in}}\sum_{i,j} \Omega_{1}(\br_j)W_{\br_i}(\br) \\
\times  \la \phi_0|   \psih^{\dagger}_{g,\br_i}(t_s) \,
\psih_{m,\br_i}(t_s) \, \psih^{\dagger}_{m,\br_j}(0) \,
\psih_{g,\br_j}(0)|\phi_0 \ra
\end{multline}
that only involves a time-dependent correlation function taken on the initial many-body state $|\phi_0\ra$.

As no atoms are initially present in the $m$ state, the initial many-body state $|\phi_0\ra$ exactly factorizes in a complex many-body state for the $g,g'$ subspace and vacuum for the $m$ one.
Assuming that the few atoms that are transferred into the $m$ state during the preparation stage do not significantly interact with the majority of atoms left in the $g,g'$ states~\cite{Stewart2008,Schunck2008,note} allows us to neglect the $m-g,g'$ interaction term of the Hamiltonian $H_{int}$ in (4) and write the time-evolution operator $U_{mb}$ in the factorized form $U_{mb}(t_s)= e^{-iH_0t_s/\hbar}e^{-iH_mt_s/\hbar}$.
As a direct consequence, the Heisenberg evolution of the $\psih_{m,\br_i}(t)$ operator is only determined by the $H_m$ part of the evolution operator, while the $\psih_{g,\br_i}(t)$ and $\psih_{g',\br_i}(t)$ operators evolve with the many-body $H_0$ Hamiltonian in the $g,g'$ space.
These simple facts allow one to rewrite the dipole expectation value in the final form:

\begin{eqnarray}
{d}({\br},t_s)
= &&-\frac{D}{\Omega_2^{\rm in}} \sum_{i,j}\Omega_{1}(\br_j))W_{\br_i}(\br)
\nonumber\\
&&\times \la\textrm{vac}| \psih_{m,\br_i}(t_s) \, \psih^{\dagger}_{m,\br_j}(0) |\textrm{vac}\ra \nonumber\\
&&\times \la \phi_0 |\psih^{\dagger}_{g,\br_i}(t_s) \,
\psih_{g,\br_j}(0)|\phi_0\ra. \label{eq:final_dipole}
\end{eqnarray}
where all the expectation values are to be evaluated on the
initial many-body state before the preparation stage, with no
occupation in the $m$ state. In particular, the $m$ state
propagator describes the free-particle evolution in the lattice
potential.

The far-field pattern in a direction $\hat{\theta}$ is
proportional to the spatial Fourier transform of ${d}({\br},t_s)$
evaluated at a wavevector $\bk$ equal to the projection of the
emission wavevector $\hat{\theta}\,\omega_1/c$ along the $xy$
plane. Here, $c$ is the velocity of light. This leads to the
following expression for the far-field emission amplitude at a
distance $\mathbf{R}=R\,\hat{\theta}$:
\begin{equation}
E^{\rm out}_{\hat{\theta}}(t_s)= \frac{C_{\bk}}{N}\,\sum_{\bq}
\,e^{-i\omega_m(\bq+\bk)t_s} \bra{\phi_0}
\psih^{\dagger}_{g,\bq}(t_s)\,\psih_{g,\bq}(0)\ket{\phi_0}
\eqname{Ek}
\end{equation}
where $\omega_m(\bq)$ is the free-particle dispersion of $m$ state atoms in the lattice potential and the coefficient $C_{\bk}$ is defined as
\begin{equation}
C_{\bk}=\frac{D\,\omega_1^2\,\Omega_{1}(\bk)\,W(\bk)}{4\pi\epsilon_0\,R\,c^2\,\Omega_2^{\rm
in}}.
\end{equation}
Invariance under translations along the plane guarantees that the
coherent emission amplitude in the $\hat{\theta}$ direction, i.e.
with an in-plane wavevector $\bk$, only depends on the incident
probe amplitude $\Omega_1(\bk)$ at the same $\bk$. Here we have
set
$\Omega_{1}(\bk)=\sum_{j}\,\Omega_{1}(\br)e^{-i\bk.\br_j}$, we
have defined $\hat{\psi}_{g,\bq}=(N)^{-1/2} \sum_{j}
\hat{\psi}_{g,\br_j}e^{-i\bq.\br_j}$, and we have used a lattice
with $N$ sites neglecting boundary effects. The factor
$W(\bk)=\int d^2\br w^*_g(\br)w_m(\br)e^{-i\bk.\br}$ is a slowly
varying envelop stemming from the tight atomic Wannier functions.

Expression \eq{Ek} relates the coherent amplitude $E^{\rm out}_{\hat{\theta}}(t_s)$ of the released light to the time-dependent one-body Green function of a generic many-body gas. It is one key result of the present paper. In the limiting case $\omega_m(\bq)=\omega_m^o$ where the $m$ atoms do not appreciably move during the time $t_s$, the far-field amplitude \eq{Ek} can be further simplified into the form
\begin{equation}
E^{\rm out}_{\hat{\theta}}(t_s)= C_{\bk} \,e^{-i \omega^o_{m}\,
t_s}  \bra{\phi_0} \psih^{\dagger}_{g,\br_i}(t_s)\,\psih_{g,\br_i}(0) \ket{\phi_0},
\eqname{Ekloc}
\end{equation}
which only involves the local value of the Green function of $g$ atoms.

Experimentally, the coherent $E^{\rm out}_{\hat{\theta}}$
amplitude can be measured by homodyne detection of the emission
with a stronger reference beam at $\omega_1$. The intensity and
phase of $E^{\rm out}_{\hat{\theta}}$ is inferred from the
amplitude and phase of the oscillations in the interference signal
as a function of the mixing phase. This procedure requires
coherence at the $g\rightarrow m$ frequency which can be easily
achieved if all $\Omega_1$, $\Omega_2^{\rm in,out}$ fields are
obtained from a single laser source.

Another quantity of interest is the {\em total} (i.e.~coherent and
incoherent) intensity pattern in either the far- or the
near-field. Differently from the coherent amplitude \eq{Ek}, these
involve higher order correlations of the many-body gas. For
instance, the near-field dipole pattern $I(\rr)= \la
\hat{d}^\dagger(\rr)\,\hat{d}(\rr) \ra$ reads:
\begin{eqnarray*}
I(\rr,t_s)&&= \frac{|D|^2}{|\Omega_2^{\rm in}|^2}
\sum_{i,j} |\Omega_1(\br_j) W_{\br_i} (\br)|^2\\
&&\times \bra{\textrm{vac}}\psih_{m,\rr_j}(0)\,\psihd_{m,\rr_i}(t_s) \psih_{m,\rr_i}(t_s)\,\psihd_{m,\rr_j}(0)|\textrm{vac}\ra \\
&&\times \bra{\phi_0}
\psih^{\dagger}_{g,\br_j}(0)\,\psih_{g,\br_i}(t_s)\,\psih^\dagger_{g,\br_i}(t_s)\,\psih_{g,\br_j}(0)
\ket{\phi_0}.
\end{eqnarray*}
For a localized beam $\Omega_1(\br_j)$, the $I(\rr)$ signal is
proportional to a fixed envelope determined by the motion of atoms
in the $m$ state times a two-body Green function of $g$ atoms. The
correlation function of the $g$ state can be understood as
measuring the density at time $t_s$ at site $r_i$ if one has
removed an atom at time $0$ at site $r_j$. This scheme looks
promising e.g.~to follow the dynamics of holes in
anti-ferromagnetic states~\cite{antiferro,Sangiovanni2006}.

\section{Application to BCS superfluid}

\label{sec:BCS}
In order to demonstrate the efficiency of the proposed detection
technique, we now calculate the signal that is expected for a
weakly attractive, unpolarized two-component Fermi gas in an
optical lattice at half filling. In particular we show how the
proposed method is able to identify a superfluid state and its
quasiparticles from the measured two-time correlation function.

In the normal state, the dispersion relation of quasiparticles is
given by the free-particle dispersion in the lattice. Here we take
the tight-binding form $\hbar\omega_{g,g',m}(\bq)=\hbar
\omega_{g,g',m}^o-2J_{g,g',m}[\cos (q_x a)+\cos (q_y a)]$. While
the $g,g'$ atoms feel the same potential, $J_{g}=J_{g'}$, the
hopping $J_m$ for the $m$ state atoms can be different. In the
following we set $\omega^o_{g'}=\omega^o_g=0$ and focus on the
case of half-filling.

 In the superfluid state, the quasiparticle
dispersion predicted by BCS theory consists of two branches
$E^{\pm}_{\bq}=\pm\sqrt{[\hbar \omega_{g}(\bq)]^2+\Delta^2}$
separated by a gap of amplitude $2\Delta$. The one-body Green
function $\mathcal{G}_{g}(\bq,t)=\la
\psih^{\dagger}_{g,\bq}(t)\,\psih_{g,\bq}(0)\rangle$ for the BCS
phase reads~\cite{Mahan}
\begin{multline*}
\mathcal{G}_{g}(\bq,t)= u^2_{\bq} f(E^+_{\bq})e^{i(\omega_{\rm
mf}+E^+_{\bq}/\hbar)t}\\+v^2_{\bq}f(E^-_{\bq})e^{i(\omega_{\rm
mf}+E^-_{\bq}/\hbar)t}.
\end{multline*}
The Bogoliubov coefficients are defined as
$u^2_{\bq},v^2_{\bq}=\frac{1}{2}\left[1 + \hbar
\omega_{g}(\bq)/E^{\pm}_{\bq}\right]$
and the Fermi distribution as $f(E)=(1+e^{E/k_BT})^{-1}$. $\omega_{\rm mf}$ is the mean-field shift \footnote{For the attractive Hubbard model with interaction strength $U$ the mean field shift is given by $\hbar\omega_{\rm mf}=U/2$.}. In what
follows, we shall focus our attention on low temperature $T$ for
which the upper branch $E^{+}_{\bq}$ is almost empty and can be
neglected. Under such an assumption the emission amplitude \eq{Ek} becomes
\begin{multline*}
E^{\rm out}_{\hat{\theta}}(t_s)= \frac{C_{\bk}}{N} \sum_{\bq}
e^{-i(\omega_m(\bq+\bk)-\omega_{\rm mf}-E^-_{\bq}/\hbar)t_s\,}
v^2_{\bq}f(E^-_{\bq}).
\end{multline*}

Its Fourier transform with respect to the storage time $t_s$ has
the form
\begin{equation*}
E^{\rm out}_{\hat{\theta}}(\omega_s)=\frac{C_{\bk}}{N}\sum_{\bq}
v^2_{\bq}\,f(E^-_{\bq})\,\delta(\omega_s-\omega_m(\bq+\bk)+\omega_{\rm
mf}+E^-_{\bq}/\hbar).
\end{equation*}
For each value $\omega_s$ of the frequency, the signal comes from the wavevectors $\bq$ which fulfill
\begin{equation}
\omega_s =\omega^o_{m}-\omega_{\rm mf}+
r\,\omega_g(\bq+\bk)-E^-_\bq/\hbar . \eqname{PolesInFrequency}
\end{equation}
In the
following we will neglect the contributions by $\omega^o_{m}-\omega_{\rm mf}$ since these can be eliminated in
the homodyne detection \footnote{For the slow outcoupling procedure, the out-coupling laser $\Omega_2^{\rm out}$ can be detuned by the mean-field shift from the resonance $m \rightarrow e$ to eliminate oscillations of frequency $\omega^o_{m}-\omega_{\rm mf}$.}.
Several regimes can be identified depending on the value of the
hopping ratio $r=J_m/J_g$.  Experimentally, the hopping amplitudes
can be varied within some range by tuning the frequency and
polarization of the lattice beams, or, if necessary, by using more complex multi-photon
transitions instead than the simple Raman scheme discussed so far.

\begin{figure}[!ht]
  \includegraphics[width=6.5cm,height=4.5cm]{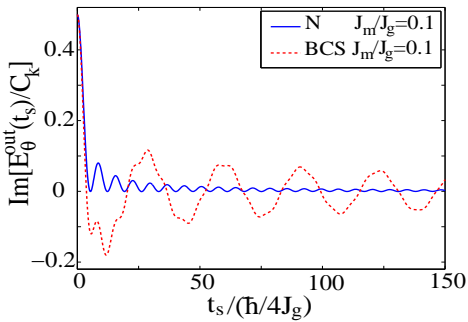}
 \caption{(Color online) Time-dependence of the emission
amplitude in the normal ($\bk=0$) direction for a normal state N
(blue solid line) and a BCS superfluid with gap $\Delta/4J_g=0.2$
(red dashed line). Hopping ratio $r=0.1$. Temperature $k_B
T=J_g/50$.}\label{fg:TimeDependent}
\end{figure}

\begin{figure}[!ht]
  \includegraphics[width=6.5cm,height=4.5cm]{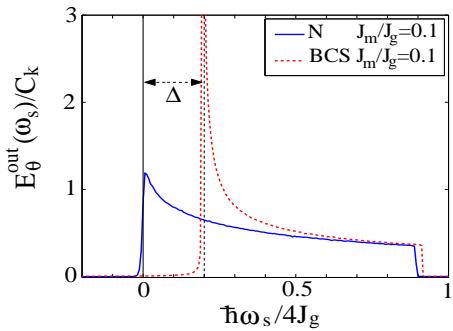}
  \caption{(Color online) Frequency-dependence of the emission
amplitude in the normal ($\bk=0$) direction for a normal state N
(blue solid line) and a BCS superfluid with gap $\Delta/4J_g=0.2$
(red dashed line). Hopping ratio $r=0.1$. Temperature $k_B
T=J_g/50$.}\label{fg:FrequencyDependentr01}
\end{figure}

\subsection{Small hopping ratio $r\ll1$}
The physics is the simplest in the $r\ll 1$ case where the atoms
in the $m$ state do not move during the experiment and the
emission amplitude is determined by the local Green function
\eq{Ekloc}. As one can see in Fig.~\ref{fg:TimeDependent}, the emission amplitude for a
superfluid state as a function of storage time $t_s$ shows a
slowly decaying oscillation at a low frequency determined by the
BCS gap $\Delta$. On top of this slow oscillation, faster and
quickly decaying oscillations are visible at frequencies on the
order of the Fermi energy (i.e. the band width $J_g$). The long
lasting, slow oscillations are a signature of the superfluid
state. They disappear in a normal state where one is left with
fast and quickly decaying oscillations.

This physics is easily understood looking at the corresponding
frequency spectra plotted in Fig.~\ref{fg:FrequencyDependentr01}. In the limiting case
$r\rightarrow 0$, the spectrum recovers the density of states for
quasi-particles. In the normal state, the spectrum has a broad
shape extending up to $ \hbar \omega_{max}=4J_g\,(1-r)$ and
showing a singularity at $\omega_s=0$ as a consequence of the
perfect nesting of the square Fermi surface at half-filling. In
the superfluid state, the dominant feature is the peak at $\hbar
\omega_s\simeq \Delta$ that limits the spectrum from below and
from which the BCS gap is immediately extracted. In this state the upper limit of the signal is shifted to
$\hbar \omega_{\max}=-4rJ_g+\sqrt{16J_g^2+\Delta^2}$.

\subsection{Equal hopping ratio $r=1$}

\begin{figure}[!ht]
  \includegraphics[width=6.5cm,height=4.5cm,clip=true]{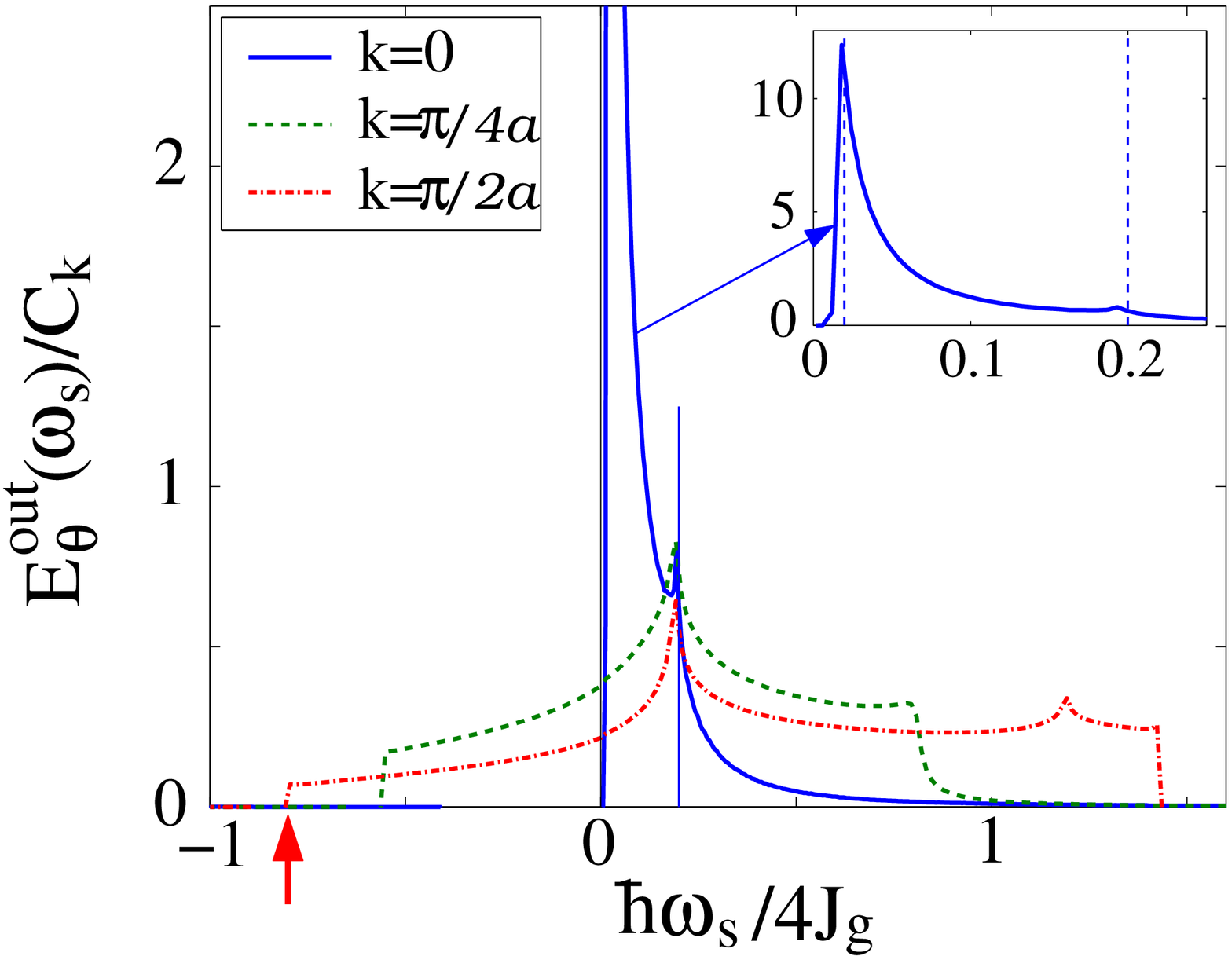}
\caption{(Color online) Frequency dependence of the emission amplitude from a BCS superfluid with
  $\Delta/4J_g=0.2$ at different angles,
$k=k_x=k_y=0,\,\pi/4a,\,\pi/2a$, with hopping ratio $r=1$.
Inset: magnified view of the $\bk=0$ curve. The arrows indicate the corresponding spectral
minimum $\omega_{\min}$. Temperature $k_B
T=J_g/50$.}\label{fg:FrequencyDependentr1}
\end{figure}

The coherent emission spectra in the case of equal hopping
amplitude $r=1$ show a rich structure that strongly depends on the
wavevector $\bk$ (Fig.~\ref{fg:FrequencyDependentr1}).
Even if the physics is somehow more involved than in the $r\ll 1$ case considered in the previous subsection, still the observed signal can be used to obtain useful information on the many-body system, e.g. its superfluid gap.

Let us first focus on  the coherent emission in the $\bk=0$
direction. At the lower boundary a large signal is
found at $\hbar \omega_{\min}\approx (\Delta^2/4J_g)/2$
for $\Delta \ll 4J_g$ (see the inset) which originates from
quasiparticles at $\bq=0$. The long tail that appears at high
frequencies past $\hbar \omega_s=\Delta$ is a direct consequence
of the smearing out of the Fermi surface on an energy scale
$\Delta$ in the BCS state.

The emission spectrum in the direction along the
diagonal of the Brillouin zone $\bk=(k,k)$ with $k=\pi/2a$ ($a$ is the
lattice constant) is characterized by two peaks and a broad
background with quite sharp edges: most visible is the strong peak
at $\hbar \omega_s=\Delta$ that originates from the divergence of
the density of states at the Fermi level in a BCS state.
This peak persists for
different values of $k$ (cf.~Fig.~\ref{fg:FrequencyDependentr1} $k=\pi/4a$) and its position can be used to
experimentally measure the amplitude $\Delta$ of the gap.

\begin{figure}[!ht]
\begin{center}  \includegraphics[width=7cm,height=5.5cm,clip=true]{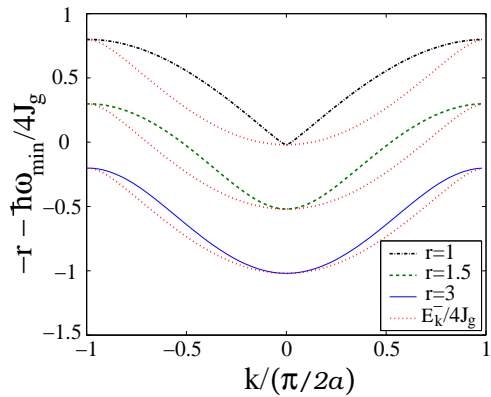}
  \caption{(Color online) $\bk$-dependence of the lower edge of the spectrum compared to the
quasiparticle dispersion $E^{-}_{\bk}$ of the BCS superfluid with
$\Delta/4J_g=0.2$. From top to bottom, hopping ratio $r=1,1.5,3$.
Curves for different $r$ are offset by $0.5$ for better
visibility. }\label{fg:MomentumDependent}
\end{center}
\end{figure}

\subsection{High hopping ratio $r\gg1$ }

\begin{figure}[!ht]
  \includegraphics[width=6.5cm,height=4.5cm,clip=true]{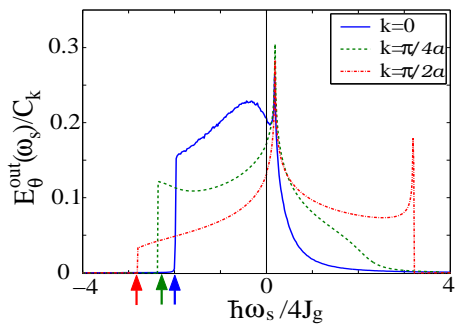}
  \caption{(Color online) Frequency dependence of the emission amplitude from a BCS superfluid with
  $\Delta/4J_g=0.2$ at different angles,
$k=k_x=k_y=0,\,\pi/4a,\,\pi/2a$. Hopping
ratio $r=3$. The arrows indicate the corresponding spectral
minimum $\omega_{\min}$. Temperature $k_B
T=J_g/50$.}\label{fg:FrequencyDependentr3}
\end{figure}

We conclude our study with a brief account of the case of a high hopping ratio $r\gg1$.
Examples of spectra for $r=3$ are plotted in Fig.~\ref{fg:FrequencyDependentr3}): in particular, they show a clear peak at $\hbar
\omega_s=\Delta$ independent of the direction of the light. This
distinctive feature allows for a direct measurement of the gap
amplitude $\Delta$.

Furthermore, the full dispersion of the BCS
quasiparticles $E^{-}_{\bq}$ can be extracted from the position of
the lower edge of the spectrum. For $r\gg 1$ the $r$-dependent
term in Eq.~\eq{PolesInFrequency} dominates and determines the
$\bq$ values that correspond to the spectral edges: the
contribution of quasi-particles with momentum $\bq=-\bk$
determines the sharp lower edge at $\hbar
\omega_{\min}=-4rJ_g-E^{-}_{-\bk}$.

The dependence of the lower
spectral edge on the emission direction $(k=k_x=k_y)$ is shown in
Fig.~\ref{fg:MomentumDependent}  for different values of $r$ and
compared to the quasi-particle dispersion. While the agreement is
limited to the special points $k=0, \pi/2a$ for $r=1$, it quickly
improves for larger $r$; a reasonably accurate image of the
quasi-particle dispersion around $k=0, \pi/2a$ is already
recovered for $r\gtrsim 3$.

\section{Conclusions and perspectives}
\label{sec:conclu}

In summary, we have proposed a novel all-optical, spatially
selective and almost non-destructive technique to probe {\em in
situ} the microscopic many-body dynamics of a gas of interacting
ultracold atoms. The technique is inspired to recent light storage experiments and is based on the creation of an atomic coherence by coherent absorption of a pump laser pulse and its later retrieval after a variable storage time: information on the many-body dynamics of the quantum gas is extracted from the amplitude and coherence properties of the retrieved light. Differently from most previous measurement schemes, the use of a spatially localized pump spot will allow to individually address the different coexisting quantum phases that can appear in a trapped system.

The efficiency of the proposed measurement scheme is tested on the specific, analytically tractable example of a two-dimensional BCS superfluid. Protocols to extract the superfluid gap and the quasi-particle dispersion are presented, which take into account some most significant difficulties that arise from the internal structure of the atoms.

As our scheme consists of the measurement of two-times correlation functions, it is expected to be of great utility in the study of the non-equilibrium dynamics of a quantum system: on one hand, its almost non-destructive nature suggests that a series of many measurements can be performed at a high repetition rate without significantly perturbing the system dynamics. On the other hand, the observed quantities play a crucial role in the characterization of relaxation dynamics \cite{CalabreseCardy2007, Cugliandolo2002}: for instance,they may serve to identify the glassiness of a  system in the presence of disorder \cite{Cugliandolo2002}.

Future work will investigate the extension of the method to more complex, three-dimensional geometries: differently from the two-dimensional geometry considered so far, this requires a careful treatment of light propagation across a bulk sample in both the excitation and the retrieval stages. Preliminary work in this direction has appeared as~\cite{Bariani2009}.

\acknowledgments We are grateful to F. Bariani, M. Capone, M. Inguscio and the Quantum Optics group of ETH Z{\"u}rich for stimulating
discussions. We acknowledge support from the 'Triangle de la Physique', ANR ('FABIOLA' and 'FAMOUS'), the DARPA-OLE program, and EPSRC (EP/G029547/1). CK would like to acknowledge the IPAM for its hospitality.


\begin{thebibliography}{99}
\bibitem{note} This requires that the timescale of the interaction between $m$ and $g,g'$ atoms is much longer than the storage time $t_s$.
\end{thebibliography}
\end{document}